\documentclass[a4paper,twocolumn]{aa}
\usepackage{amsmath,amssymb}
\usepackage[T1]{fontenc}
\usepackage{graphicx,subfigure}
\usepackage[percent]{overpic}
\usepackage{natbib}
\bibpunct{(}{)}{;}{a}{}{,} 

\renewcommand{\l}{\ell}
\newcommand{\bb}{\boldsymbol}
\newcommand{\T}{\dagger}

\begin{document}

\title{NIFT{\Large Y}\thanks{\textsc{NIFTy} homepage \url{http://www.mpa-garching.mpg.de/ift/nifty/}; Excerpts of this paper are part of the \textsc{NIFTy} source code and documentation.} -- Numerical Information Field Theory}
\subtitle{a versatile P{\large YTHON} library for signal inference}

\titlerunning{\textsc{NIFTy} -- Numerical Information Field Theory}

\author{
    Marco~Selig\inst{\ref{inst1}}\and
    Michael~R.~Bell\inst{\ref{inst1}}\and
    Henrik~Junklewitz\inst{\ref{inst1}}\and
    Niels~Oppermann\inst{\ref{inst1}}\and \\
    Martin~Reinecke\inst{\ref{inst1}}\and
    Maksim~Greiner\inst{\ref{inst1}}\inst{\ref{inst2}}\and
    Carlos~Pachajoa\inst{\ref{inst1}}\inst{\ref{inst3}}\and
    Torsten~A.~En{\ss}lin\inst{\ref{inst1}}
}

\institute{
    Max Planck Institute f\"ur Astrophysik (Karl-Schwarzschild-Stra{\ss}e~1, D-85748~Garching, Germany)\label{inst1} \and
    Ludwig-Maximilians-Universit\"at M\"unchen (Geschwister-Scholl-Platz~1, D-80539~M\"unchen, Germany) \label{inst2} \and
    Technische Universit\"at M\"unchen (Arcisstra{\ss}e~21, D-80333~M\"unchen, Germany) \label{inst3}
}

\date{Received 05 Feb. 2013 / Accepted 12 Apr. 2013}

\abstract{
    \textsc{NIFTy}, ``Numerical Information Field Theory'', is a software package designed to enable the development of signal inference algorithms that operate regardless of the underlying spatial grid and its resolution. Its object-oriented framework is written in \textsc{Python}, although it accesses libraries written in \textsc{Cython}, C++, and C for efficiency.
    \textsc{NIFTy} offers a toolkit that abstracts discretized representations of continuous spaces, fields in these spaces, and operators acting on fields into classes. Thereby, the correct normalization of operations on fields is taken care of automatically without concerning the user. This allows for an abstract formulation and programming of inference algorithms, including those derived within information field theory. Thus, \textsc{NIFTy} permits its user to rapidly prototype algorithms in 1D, and then apply the developed code in higher-dimensional settings of real world problems. The set of spaces on which \textsc{NIFTy} operates comprises point sets, $n$-dimensional regular grids, spherical spaces, their harmonic counterparts, and product spaces constructed as combinations of those.
    The functionality and diversity of the package is demonstrated by a Wiener filter code example that successfully runs without modification regardless of the space on which the inference problem is defined.
}


\maketitle

\section{Introduction}

    In many signal inference problems, one tries to reconstruct a continuous signal field from a finite set of experimental data. The finiteness of data sets is due to their incompleteness, resolution, and the sheer duration of the experiment. A further complication is the inevitability of experimental noise, which can arise from various origins. Numerous methodological approaches to such inference problems are known in modern information theory founded by \citet{C46,S48,W49}.

    Signal inference methods are commonly formulated in an abstract, mathematical way to be applicable in various scenarios; i.e., the method itself is independent, or at least partially independent, of resolution, geometry, physical size, or even dimensionality of the inference problem. It then is up to the user to apply the appropriate method correctly to the problem at hand.

    In practice, signal inference problems are solved numerically, rather than analytically. Numerical algorithms should try to preserve as much of the universality of the underlying inference method as possible, given the limitations of a computer environment, so that the code is reuseable. For example, an inference algorithm developed in astrophysics that reconstructs the photon flux on the sky from high energy photon counts might also serve the purpose of reconstructing two- or three-dimensional medical images obtained from tomographical X-rays. The desire for multi-purpose, problem-independent inference algorithms is one motivation for the \textsc{NIFTy} package presented here. Another is to facilitate the implementation of problem specific algorithms by providing many of the essential operations in a convenient way.

    \textsc{NIFTy} stands for ``Numerical Information Field Theory''. It is a software package written in \textsc{Python}\footnote{\textsc{Python} homepage \url{http://www.python.org/}}\footnote{\textsc{NIFTy} is written in \textsc{Python} 2 which is supported by all platforms and compatible to existing third party packages. A \textsc{Python} 3 compliant version is left for a future upgrade.}, however, it also incorporates \textsc{Cython}\footnote{\textsc{Cython} homepage \url{http://cython.org/}} \citep{B+09,S09}, C++, and C libraries for efficient computing.

    The purpose of the \textsc{NIFTy} library is to provide a toolkit that enables users to implement their algorithms as abstractly as they are formulated mathematically. \textsc{NIFTy}'s field of application is kept broad and not bound to one specific methodology. The implementation of maximum entropy \citep{J57,J89}, likelihood-free, maximum likelihood, or full Bayesian inference methods \citep{B63,L,C46} are feasible, as well as the implementation of posterior sampling procedures based on Markov chain Monte Carlo procedures \citep{MU49,M+53}.

    Although \textsc{NIFTy} is versatile, the original intention was the implementation of inference algorithms that are formulated methodically in the language of information field theory\footnote{IFT homepage \url{http://www.mpa-garching.mpg.de/ift/}} (IFT). The idea of IFT is to apply information theory to the problem of signal field inference, where ``field'' is the physicist's term for a continuous function over a continuous space. The recovery of a field that has an infinite number of degrees of freedom from finite data can be achieved by exploiting the spatial continuity of fields and their internal correlation structures. The framework of IFT is detailed in the work by \citet{EFK09} where the focus lies on a field theoretical approach to inference problems based on Feynman diagrams. An alternative approach using entropic matching based on the formalism of the Gibbs free energy can be found in the work by \citet{EW10}. IFT based methods have been developed to reconstruct signal fields without \emph{a priori} knowledge of signal and noise correlation structures \citep{EF11,ORE11}. Furthermore, IFT has been applied to a number of problems in astrophysics, namely to recover the large scale structure in the cosmic matter distribution using galaxy counts \citep{KJ+09,JKWE10,JK10,JKLE10,WE10}, and to reconstruct the Faraday rotation of the Milky Way \citep{O+11}. A more abstract application has been shown to improve stochastic estimates such as the calculation of matrix diagonals by sample averages \citep{SOE12}.

    One natural requirement of signal inference algorithms is their independence of the choice of a particular grid and a specific resolution, so that the code is easily transferable to problems that are similar in terms of the necessary inference methodology but might differ in terms of geometry or dimensionality. In response to this requirement, \textsc{NIFTy} comprises several commonly used pixelization schemes and their corresponding harmonic bases in an object-oriented framework. Furthermore, \textsc{NIFTy} preserves the continuous limit by taking care of the correct normalization of operations like scalar products, matrix-vector multiplications, and grid transformations; i.e., all operations involving position integrals over continuous domains.

    The remainder of this paper is structured as follows. In Sec.~\ref{sec:inference} an introduction to signal inference is given, with the focus on the representation of continuous information fields in the discrete computer environment. Sec.~\ref{sec:overview} provides an overview of the class hierarchy and features of the \textsc{NIFTy} package. The implementation of a Wiener filter algorithm demonstrates the basic functionality of \textsc{NIFTy} in Sec.~\ref{sec:demonstration}. We conclude in Sec.~\ref{sec:conclusion}.

\section{Concepts of Signal Inference}
\label{sec:inference}

\subsection{Fundamental Problem}

    Many signal inference problems can be reduced to a single model equation,
    \begin{align}
        \label{eq:abstractmodel}
        \bb{d} &= f(\bb{s},\dots)
        ,
    \end{align}
    where the data set $\bb{d}$ is the outcome of some function $f$ being applied to a set of unknowns.\footnote{An alternative notation commonly found in the literature is $\bb{y} = f[\bb{x}]$. We do not use this notation in order to avoid confusion with coordinate variables, which in physics are commonly denoted by $x$ and $y$.} Some of the unknowns are of interest and form the signal $\bb{s}$, whereas the remaining are considered as nuisance parameters. The goal of any inference algorithm is to obtain an approximation for the signal that is ``best'' supported by the data. Which criteria define this ``best'' is answered differently by different inference methodologies.

    There is in general no chance of a direct inversion of Eq.~\eqref{eq:abstractmodel}. Any realistic measurement involves random processes summarized as noise and, even for deterministic or noiseless measurement processes, the number of degrees of freedom of a signal typically outnumbers those of a finite data set measured from it, because the signal of interest might be a continuous field; e.g., some physical flux or density distribution.

    In order to clarify the concept of measuring a continuous signal field, let us consider a linear measurement by some response $\bb{R}$ with additive and signal independent noise $\bb{n}$,
    \begin{align}
        \label{eq:linearmodel}
        \bb{d} &= \bb{R}\bb{s} + \bb{n}
        ,
    \end{align}
    which reads for the individual data points,
    \begin{align}
        \label{eq:linearmodelindexed}
        d_i &= \int_{\Omega}\mathrm{d}x \; R_i(x)s(x) + n_i
        .
    \end{align}
    Here we introduced the discrete index $i \in \{1,\dots,N\} \subset \mathbb{N}$ and the continuous position $x \in \Omega$ of some abstract position space $\Omega$. For example, in the context of image reconstruction, $i$ could label the $N$ image pixels and $x$ would describe real space positions.

    The model given by Eq.~\eqref{eq:linearmodel} already poses a full inference problem since it involves an additive random process and a non-invertible signal response. As a consequence, there are many possible field configurations in the signal phase space that could explain a given data set. The approach used to single out the ``best'' estimate of the signal field from the data at hand is up to the choice of inference methodology. However, the implementation of any derived inference algorithm needs a proper discretization scheme for the fields defined on $\Omega$.
    Since one might want to extend the domain of application of a successful algorithm, it is worthwhile to keep the implementation flexible with respect to the characteristics of $\Omega$.

    \begin{table*}[!t]
        \caption{Overview of derivatives of the \textsc{NIFTy} space class, the corresponding grids, and conjugate space classes.}
        \centering
        \begin{tabular}{|rll|}
            \hline
            \multicolumn{1}{|c}{\textsc{NIFTy} subclass} & \multicolumn{1}{c}{corresponding grid} & \multicolumn{1}{c|}{conjugate space class} \\
            \hline
            \hline
            \texttt{point\_space} & unstructured list of points & (none) \\
            \texttt{rg\_space} & $n$-dimensional regular Euclidean grid over $\mathcal{T}^n$ & \texttt{rg\_space} \\
            \texttt{lm\_space} & spherical harmonics & \texttt{gl\_space} or \texttt{hp\_space} \\
            \texttt{gl\_space} & Gauss-Legendre grid on the $\mathcal{S}^2$ sphere & \texttt{lm\_space} \\
            \texttt{hp\_space} & \textsc{HEALPix} grid on the $\mathcal{S}^2$ sphere & \texttt{lm\_space} \\
            \texttt{nested\_space} & (arbitrary product of grids) & (partial conjugation) \\
            \hline
        \end{tabular}
        \label{tab:spaces}
    \end{table*}

\subsection{Discretized Continuum}
\label{sec:discretization}

    The representation of fields that are mathematically defined on a continuous space in a finite computer environment is a common necessity. The goal hereby is to preserve the continuum limit in the calculus in order to ensure a resolution independent discretization.

    Any partition of the continuous position space $\Omega$ (with volume $V$) into a set of $Q$ disjoint, proper subsets $\Omega_q$ (with volumes $V_q$) defines a pixelization,
    \begin{align}
        \Omega &= \dot{\bigcup_q} \; \Omega_q \qquad \mathrm{with} \quad q \in \{1,\dots,Q\} \subset \mathbb{N}
        , \\
        \label{eq:vol}
        V &= \int_\Omega \mathrm{d}x = \sum_{q=1}^Q \int_{\Omega_q} \mathrm{d}x = \sum_{q=1}^Q V_q
        .
    \end{align}
    Here the number $Q$ characterizes the resolution of the pixelization, and the continuum limit is described by $Q \rightarrow \infty$ and $V_q \rightarrow 0$ for all $q \in \{1,\dots,Q\}$ simultaneously. Moreover, Eq.~\eqref{eq:vol} defines a discretization of continuous integrals, $\int_\Omega \mathrm{d}x \mapsto \sum_q V_q$.

    Any valid discretization scheme for a field $\bb{s}$ can be described by a mapping,
    \begin{align}
        s(x \in \Omega_q) \mapsto s_q = \int_{\Omega_q} \mathrm{d}x \; w_q(x) s(x)
        ,
    \end{align}
    if the weighting function $w_q(x)$ is chosen appropriately. In order for the discretized version of the field to converge to the actual field in the continuum limit, the weighting functions need to be normalized in each subset; i.e., $\forall q: \int_{\Omega_q} \mathrm{d}x \; w_q(x) = 1$. Choosing such a weighting function that is constant with respect to $x$ yields
    \begin{align}
        \label{eq:discretization}
        s_q = \frac{\int_{\Omega_q} \mathrm{d}x \; s(x)}{\int_{\Omega_q} \mathrm{d}x} = \left< s(x) \right>_{\Omega_q}
        ,
    \end{align}
    which corresponds to a discretization of the field by spatial averaging. Another common and equally valid choice is $w_q(x) = \delta(x-x_q)$, which distinguishes some position $x_q \in \Omega_q$, and evaluates the continuous field at this position,
    \begin{align}
        s_q = \int_{\Omega_q} \mathrm{d}x \; \delta(x-x_q) s(x) = s(x_q)
        .
    \end{align}
    In practice, one often makes use of the spatially averaged pixel position, $x_q = \left< x \right>_{\Omega_q}$; cf. Eq.~\eqref{eq:discretization}. If the resolution is high enough to resolve all features of the signal field $\bb{s}$, both of these discretization schemes approximate each other, $\left< s(x) \right>_{\Omega_q} \approx s(\left< x \right>_{\Omega_q})$, since they approximate the continuum limit by construction.\footnote{The approximation of $\left< s(x) \right>_{\Omega_q} \approx s(x_q \in \Omega_q)$ marks a resolution threshold beyond which further refinement of the discretization reveals no new features; i.e., no new information content of the field $\bb{s}$.}

    All operations involving position integrals can be normalized in accordance with Eqs. \eqref{eq:vol} and \eqref{eq:discretization}. For example, the scalar product between two fields $\bb{s}$ and $\bb{u}$ is defined as
    \begin{align}
        \label{eq:dot}
        \bb{s}^\T \bb{u} = \int_\Omega \mathrm{d}x \; s^*(x) u(x) \approx \sum_{q=1}^Q V_q^{\phantom{*}} s_q^* u_q^{\phantom{*}}
        ,
    \end{align}
    where $\T$ denotes adjunction and $*$ complex conjugation. Since the approximation in Eq.~\eqref{eq:dot} becomes an equality in the continuum limit, the scalar product is independent of the pixelization scheme and resolution, if the latter is sufficiently high.

    The above line of argumentation analogously applies to the discretization of operators. For a linear operator $\bb{A}$ acting on some field $\bb{s}$ as $\bb{A} \bb{s} = \int_\Omega \mathrm{d}y A(x,y) s(y)$, a matrix representation discretized in analogy to Eq.~\eqref{eq:discretization} is given by
    \begin{align}
        A(x \in \Omega_p, y \in \Omega_q) \mapsto A_{pq} &= \frac{\iint_{\Omega_p \Omega_q} \mathrm{d}x \, \mathrm{d}y \; A(x,y)}{\iint_{\Omega_p \Omega_q} \mathrm{d}x \, \mathrm{d}y}
        \nonumber \\
        &= \big< \big< A(x,y) \big>_{\Omega_p} \big>_{\Omega_q}
        .
    \end{align}
    Consequential subtleties regarding operators are addressed in App.~\ref{sec:matrices}.

    The proper discretization of spaces, fields, and operators, as well as the normalization of position integrals, is essential for the conservation of the continuum limit. Their consistent implementation in \textsc{NIFTy} allows a pixelization independent coding of algorithms.

\section{Class and Feature Overview}
\label{sec:overview}

    The \textsc{NIFTy} library features three main classes: spaces that represent certain grids, fields that are defined on spaces, and operators that apply to fields. In the following, we will introduce the concept of these classes and comment on further \textsc{NIFTy} features such as operator probing.

\subsection{Spaces}

    The \texttt{space} class is an abstract class from which all other specific space subclasses are derived. Each subclass represents a grid type and replaces some of the inherited methods with its own methods that are unique to the respective grid. This framework ensures an abstract handling of spaces independent of the underlying geometrical grid and the grid's resolution.

    An instance of a space subclass represents a geometrical space approximated by a specific grid in the computer environment. Therefore, each subclass needs to capture all structural and dimensional specifics of the grid and all computationally relevant quantities such as the data type of associated field values. These parameters are stored as properties of an instance of the class at its initialization, and they do not need to be accessed explicitly by the user thereafter. This prevents the writing of grid or resolution dependent code.

    Spatial symmetries of a system can be exploited by corresponding coordinate transformations. Often, transformations from one basis to its harmonic counterpart can greatly reduce the computational complexity of algorithms. The harmonic basis is defined by the eigenbasis of the Laplace operator; e.g., for a flat position space it is the Fourier basis.\footnote{The covariance of a Gaussian random field that is statistically homogeneous in position space becomes diagonal in the harmonic basis.} This conjugation of bases is implemented in \textsc{NIFTy} by distinguishing conjugate space classes, which can be obtained by the instance method \texttt{get\_codomain} (and checked for by \texttt{check\_codomain}). Moreover, transformations between conjugate spaces are performed automatically if required.

    Thus far, \textsc{NIFTy} has six classes that are derived from the abstract space class. These subclasses are described here, and an overview can be found in Tab.~\ref{tab:spaces}.

    \begin{itemize}
        \item[$\bullet$]
            The \texttt{point\_space} class merely embodies a geometrically unstructured list of points. This simplest possible kind of grid has only one parameter, the total number of points. This space is thought to be used as a default data space and neither has a conjugate space nor matches any continuum limit.
        \item[$\bullet$]
            The \texttt{rg\_space} class comprises all regular Euclidean grids of arbitrary dimension and periodic boundary conditions. Such a grid is described by the number of grid points per dimension, the edge lengths of one $n$-dimensional pixel and a few flags specifying the origin of ordinates, internal symmetry, and basis type; i.e., whether the grid represents a position or Fourier basis. The conjugate space of a \texttt{rg\_space} is another \texttt{rg\_space} that is obtained by a fast Fourier transformation of the position basis yielding a Fourier basis or vice versa by an inverse fast Fourier transformation.
        \item[$\bullet$]
            The spherical harmonics basis is represented by the \texttt{lm\_space} class which is defined by the maximum of the angular and azimuthal quantum numbers, $\l$ and $m$, where $m_\mathrm{max} \leq \l_\mathrm{max}$ and equality is the default. It serves as the harmonic basis for the instance of both the \texttt{gl\_space} and the \texttt{hp\_space} class.
        \item[$\bullet$]
            The \texttt{gl\_space} class describes a Gauss-Legendre grid on an $\mathcal{S}^2$ sphere, where the pixels are centered at the roots of Gauss-Legendre polynomials. A grid representation is defined by the number of latitudinal and longitudinal bins, $n_\mathrm{lat}$ and $n_\mathrm{lon}$.
        \item[$\bullet$]
            The hierarchical equal area isolatitude pixelization of an $\mathcal{S}^2$ sphere (abbreviated as \textsc{HEALPix}\footnote{\textsc{HEALPix} homepage \url{http://sourceforge.net/projects/healpix/}}) is represented by the \texttt{hp\_space} class. The grid is characterized by twelve basis pixels and the $n_\mathrm{side}$ parameter that specifies how often each of them is quartered.
        \item[$\bullet$]
            The \texttt{nested\_space} class is designed to comprise all possible product spaces constructed out of those described above. Therefore, it is defined by an ordered list of space instances that are meant to be multiplied by an outer product. Conjugation of this space is conducted separately for each subspace.

            For example, a 2D regular grid can be cast to a nesting of two 1D regular grids that would then allow for separate Fourier transformations along one of the two axes.
    \end{itemize}

\subsection{Fields}

    \begin{table*}[!t]
        \caption{Selection of instance methods of the \textsc{NIFTy} field class.}
        \centering
        \begin{tabular*}{\textwidth}{@{\extracolsep{\fill}}|rl|}
            \hline
            \multicolumn{1}{|c}{method name} & \multicolumn{1}{c|}{description} \\
            \hline
            \hline
            \texttt{cast\_domain} & alters the field's domain without altering the field values or the codomain. \\
            \texttt{conjugate} & complex conjugates the field values. \\
            \texttt{dot} & applies the scalar product between two fields, returns a scalar. \\
            \texttt{tensor\_dot} & applies a tensor product between two fields, returns a field defined in the product space. \\
            \texttt{pseudo\_dot} & applies a scalar product between two fields on a certain subspace of a product space, returns a \\ & $\quad$ scalar or a field, depending on the subspace. \\
            \texttt{dim} & returns the dimensionality of the field. \\
            \texttt{norm} & returns the $L^2$-norm of the field. \\
            \texttt{plot} & draws a figure illustrating the field. \\
            \texttt{set\_target} & alters the field's codomain without altering the domain or the field values. \\
            \texttt{set\_val} & alters the field values without altering the domain or codomain. \\
            \texttt{smooth} & smoothes the field values in position space by convolution with a Gaussian kernel. \\
            \texttt{transform} & applies a transformation from the field's domain to some codomain. \\
            \texttt{weight} & multiplies the field with the grid's volume factors (to a given power). \\
            (and more) & \\
            \hline
        \end{tabular*}
        \label{tab:field}
    \end{table*}

    The second fundamental \textsc{NIFTy} class is the \texttt{field} class whose purpose is to represent discretized fields. Each field instance has not only a property referencing an array of field values, but also \texttt{domain} and \texttt{target} properties. The domain needs to be stated during initialization to clarify in which space the field is defined. Optionally, one can specify a target space as codomain for transformations; by default the conjugate space of the domain is used as the target space.

    In this way, a field is not only implemented as a simple array, but as a class instance carrying an array of values and information about the geometry of its domain. Calling field methods then invokes the appropriate methods of the respective space without any additional input from the user. For example, the scalar product, computed by \texttt{field.dot}, applies the correct weighting with volume factors as addressed in Sec.~\ref{sec:discretization} and performs basis transformations if the two fields to be scalar-multiplied are defined on different but conjugate domains.\footnote{Since the scalar product by discrete summation approximates the integration in its continuum limit, it does not matter in which basis it is computed.} The same is true for all other methods applicable to fields; see Tab.~\ref{tab:field} for a selection of those instance methods.

    Furthermore, \textsc{NIFTy} overloads standard operations for fields in order to support a transparent implementation of algorithms. Thus, it is possible to combine field instances by $+,-,\ast,/,\dots$ and to apply trigonometric, exponential, and logarithmic functions componentwise to fields in their current domain.

\subsection{Operators}

    \begin{table*}[!t]
        \caption{Overview of derivatives of the \textsc{NIFTy} operator class.}
        \centering
        \begin{tabular*}{\textwidth}{@{\extracolsep{\fill}}|ll|}
            \hline
            \multicolumn{1}{|c}{\textsc{NIFTy} subclass} & \multicolumn{1}{c|}{description} \\
            \hline
            \hline
            \texttt{operator} & \\
            $\hookrightarrow$ \texttt{diagonal\_operator} & representing diagonal matrices in a specified space. \\
            $\phantom{\hookrightarrow} \hookrightarrow$ \texttt{power\_operator} & representing covariance matrices that are defined by a power spectrum of a statistically \\ & $\quad$ homogeneous and isotropic random field. \\
            $\hookrightarrow$ \texttt{projection\_operator} & representing projections onto subsets of the basis of a specified space. \\
            $\hookrightarrow$ \texttt{vecvec\_operator} & representing matrices of the form $\bb{A} = \bb{a}\bb{a}^\T$, where $\bb{a}$ is a field. \\
            $\hookrightarrow$ \texttt{response\_operator} & representing an exemplary response including a convolution, masking and projection. \\
            \hline
        \end{tabular*}
        \label{tab:operators}
    \end{table*}

    Up to this point, we abstracted fields and their domains leaving us with a toolkit capable of performing normalizations, field-field operations, and harmonic transformations. Now, we introduce the generic \texttt{operator} class from which other, concrete operators can be derived.

    In order to have a blueprint for operators capable of handling fields, any application of operators is split into a general and a concrete part. The general part comprises the correct involvement of normalizations and transformations, necessary for any operator type, while the concrete part is unique for each operator subclass. In analogy to the field class, any operator instance has a set of properties that specify its domain and target as well as some additional flags.

    For example, the application of an operator $\bb{A}$ to a field $\bb{s}$ is coded as \texttt{A(s)}, or equivalently \texttt{A.times(s)}. The instance method \texttt{times} then invokes \texttt{\_briefing}, \texttt{\_multiply} and \texttt{\_debriefing} consecutively. The briefing and debriefing are generic methods in which in- and output are checked; e.g., the input field might be transformed automatically during the briefing to match the operators domain. The \texttt{\_multiply} method, being the concrete part, is the only contribution coded by the user. This can be done both explicitly by multiplication with a complete matrix or implicitly by a computer routine.

    There are a number of basic operators that often appear in inference algorithms and are therefore preimplemented in \textsc{NIFTy}. An overview of preimplemented derivatives of the \texttt{operator} class can be found in Tab.~\ref{tab:operators}.

\subsection{Operator Probing}

    While properties of a linear operator, such as its diagonal, are directely accessible in case of an explicitly given matrix, there is no direct approach for  implicitly stated operators. Even a brute force approach to calculate the diagonal elements one by one may be prohibited in such cases by the high dimensionality of the problem.

    That is why the \textsc{NIFTy} library features a generic \texttt{probing} class. The basic idea of probing \citep{H89} is to approximate properties of implicit operators that are only accessible at a high computational expense by using sample averages. Individual samples are generated by a random process constructed to project the quantity of interest. For example, an approximation of the trace or diagonal of a linear operator $\bb{A}$ (neglecting the discretization subtleties) can be obtained by
    \begin{align}
        \mathrm{tr}[\bb{A}] &\approx \left< \bb{\xi}^\T \bb{A} \bb{\xi} \right>_{\{\bb{\xi}\}}
        = \sum_{pq} A_{pq} \left< \xi_p \xi_q \right>_{\{\bb{\xi}\}} \rightarrow \sum_{p} A_{pp}
        , \\
        \label{eq:diag}
        \Big( \mathrm{diag}[\bb{A}] \Big)_p &\approx \left( \left< \bb{\xi} \ast \bb{A} \bb{\xi} \right>_{\{\bb{\xi}\}} \right)_p
        = \sum_q A_{pq} \left< \xi_p \xi_q \right>_{\{\bb{\xi}\}} \rightarrow A_{pp}
        ,
    \end{align}
    where $\left<\,\cdot\,\right>_{\{\bb{\xi}\}}$ is the sample average of a sample of random fields $\bb{\xi}$ with the property $\left< \xi_p \xi_q \right>_{\{\bb{\xi}\}} \rightarrow \delta_{pq}$ for $|\{\bb{\xi}\}| \rightarrow \infty$ and $\ast$ denotes componentwise multiplication, cf. \citep[and references therein]{SOE12}. One of many possible choices for the random values of $\bb{\xi}$ are equally probable values of $\pm 1$ as originally suggested by \citet{H89}. Since the residual error of the approximation decreases with the number of used samples, one obtains the exact result in the limit of infinitely many samples. In practice, however, one has to find a tradeoff between acceptable numerical accuracy and affordable computational cost.

    The \textsc{NIFTy} probing class allows for the implementation of arbitrary probing schemes. Because each sample can be computed independently, all probing operations take advantage of parallel processing for reasons of efficiency, by default. There are two derivatives of the probing class implemented in \textsc{NIFTy}, the \texttt{trace\_probing} and \texttt{diagonal\_probing} subclasses, which enable the probing of traces and diagonals of operators, respectively.

    An extension to improve the probing of continuous operators by exploiting their internal correlation structure as suggested in the work by \citet{SOE12} is planned for a future version of \textsc{NIFTy}.

\subsection{Parallelization}

    The parallelization of computational tasks is supported. \textsc{NIFTy} itself uses a shared memory parallelization provided by the \textsc{Python} standard library \texttt{multiprocessing}\footnote{\textsc{Python} documentation \url{http://docs.python.org/2/library/multiprocessing.html}} for probing. If parallelization within \text{NIFTy} is not desired or needed, it can be turned off by the global setting flag \texttt{about.multiprocessing}.

    Nested parallelization is not supported by \textsc{Python}; i.e., the user has to decide between the useage of parallel processing either within \textsc{NIFTy} or within dependent libraries such as \textsc{HEALPix}.


\section{Demonstration}
\label{sec:demonstration}

    \begin{figure*}[!t]
        \centering
        \begin{tabular}{ccc}
            \begin{overpic} [scale=0.3]{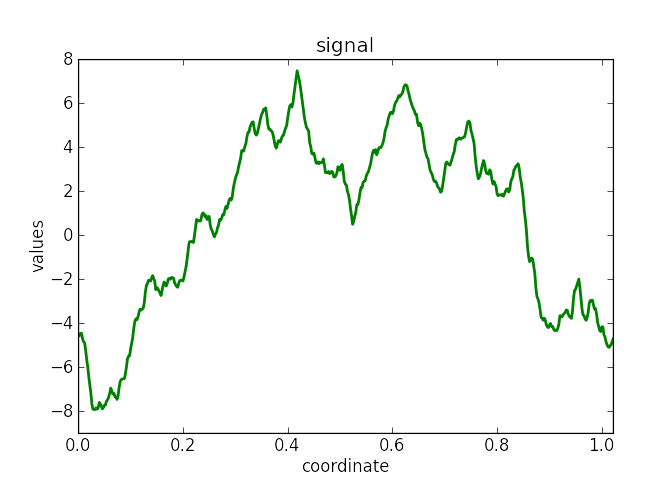} \put(-5.5,64){(a)} \end{overpic} &
            \begin{overpic} [scale=0.3]{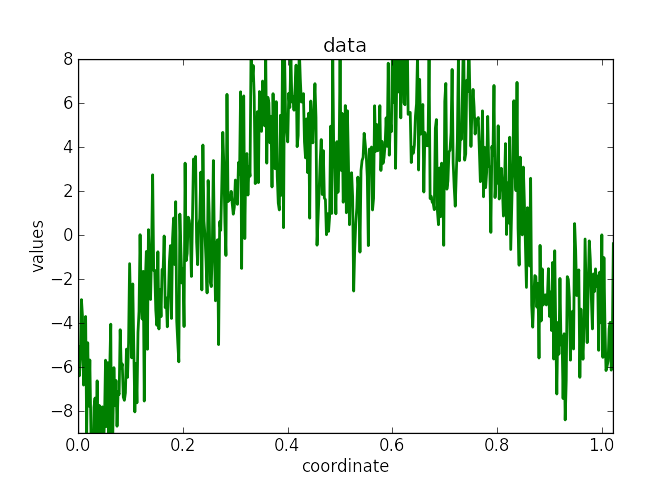} \put(-5.5,64){(b)} \end{overpic} &
            \begin{overpic} [scale=0.3]{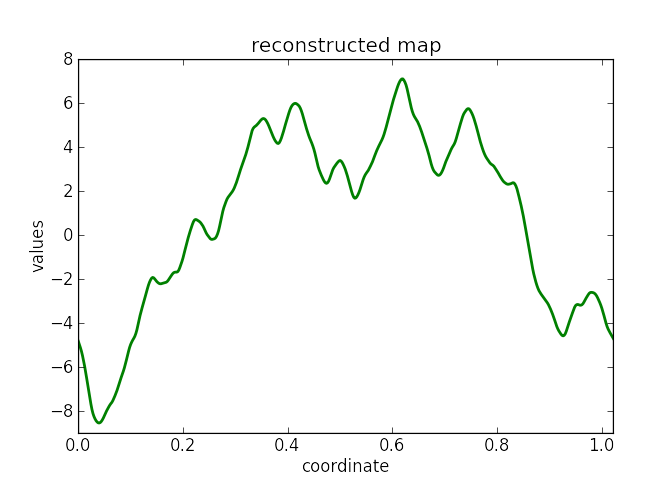} \put(-5.5,64){(c)} \end{overpic} \\
            \begin{overpic} [scale=0.3]{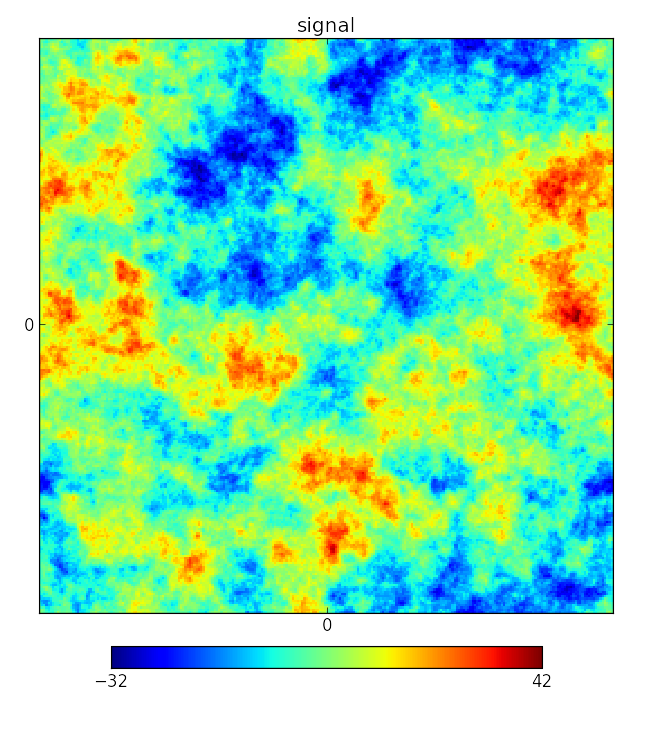} \put(-5,92){(d)} \end{overpic} &
            \begin{overpic} [scale=0.3]{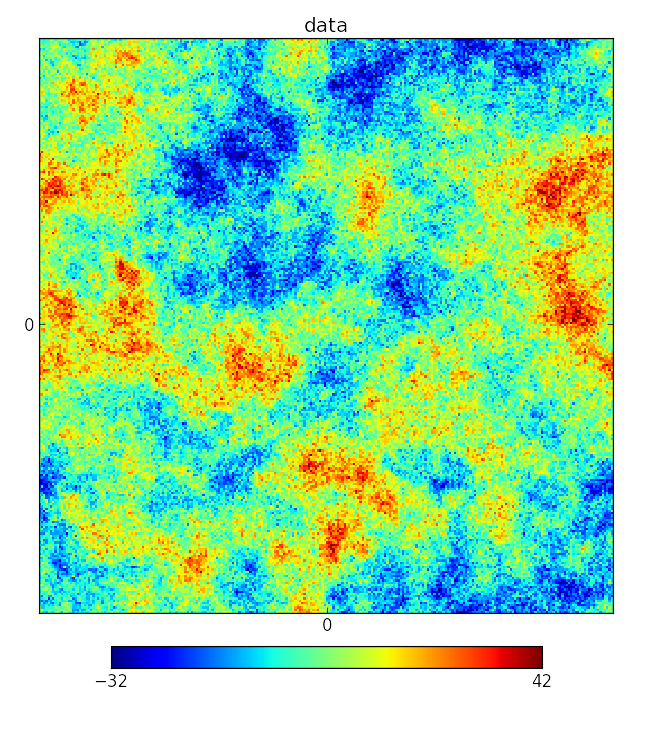} \put(-5,92){(e)} \end{overpic} &
            \begin{overpic} [scale=0.3]{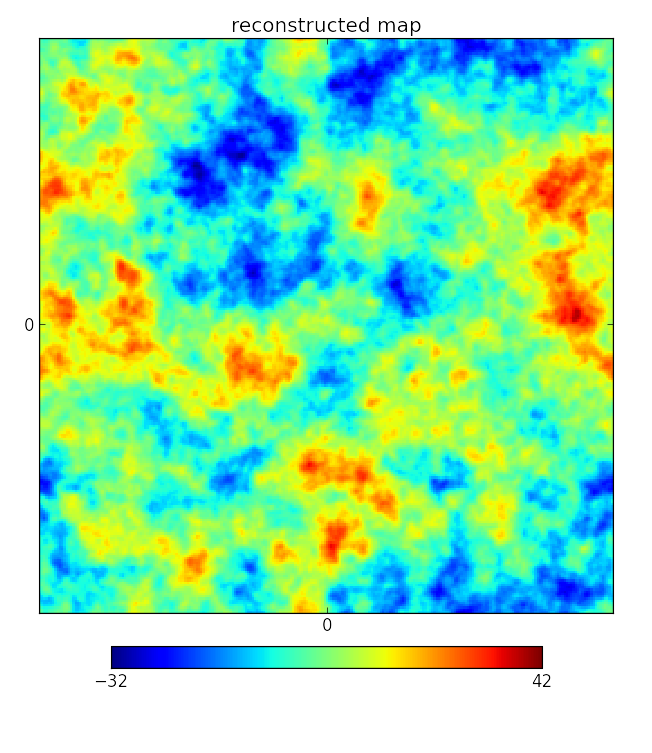} \put(-5,92){(f)} \end{overpic} \\
            \begin{overpic} [scale=0.24]{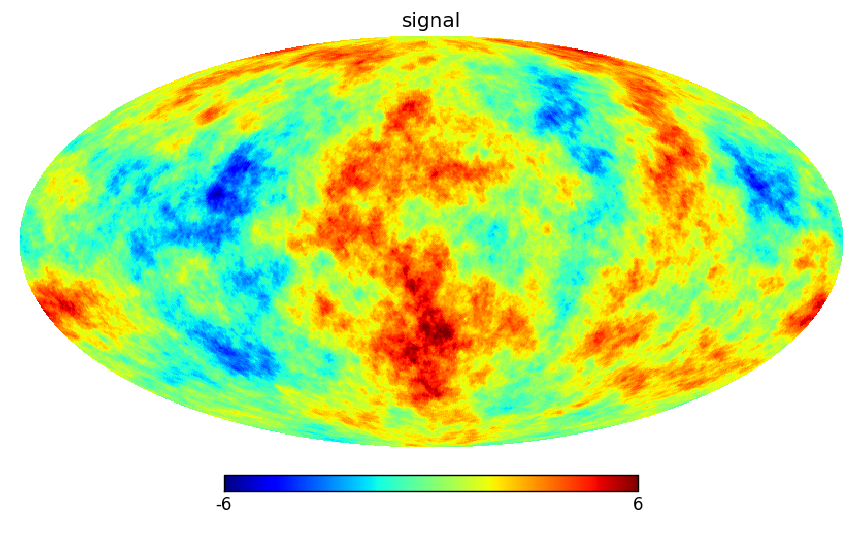} \put(-2,57){(g)} \end{overpic} &
            \begin{overpic} [scale=0.24]{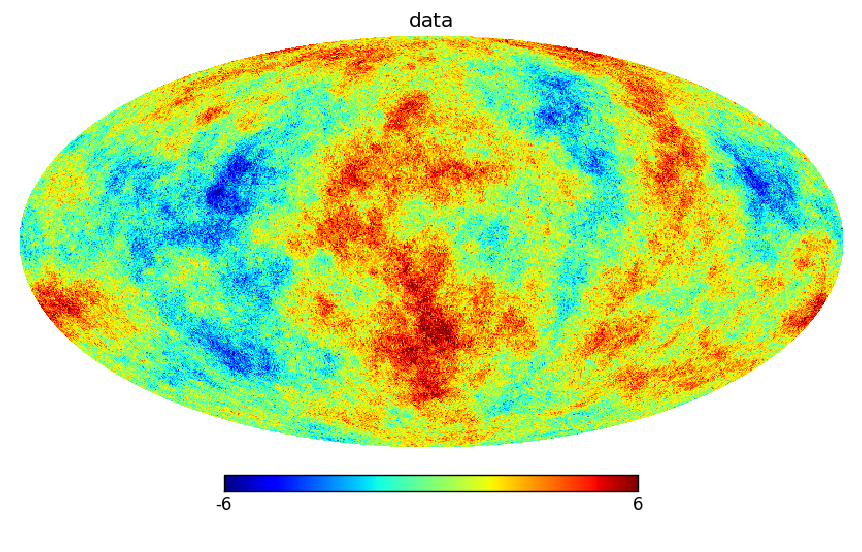} \put(-2,57){(h)} \end{overpic} &
            \begin{overpic} [scale=0.24]{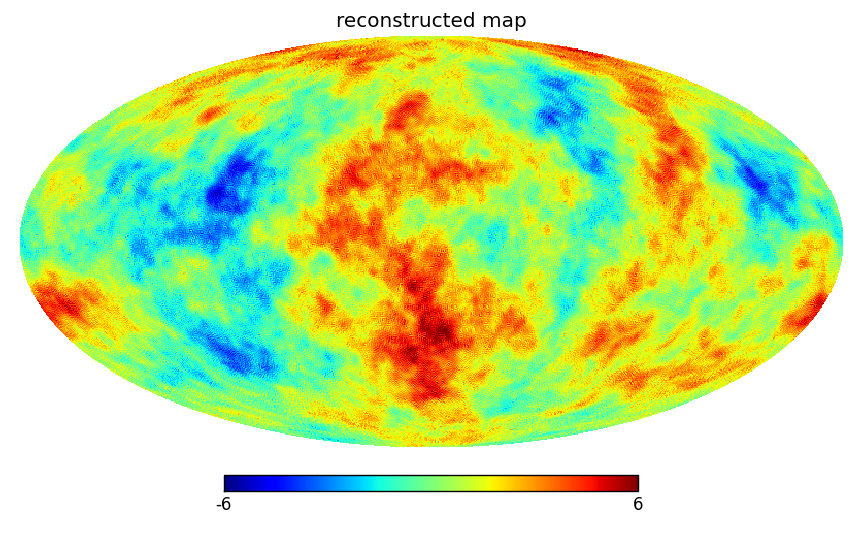} \put(-2,57){(i)} \end{overpic} \\
        \end{tabular}
        \flushleft
        \caption{Illustration of the Wiener filter code example showing (left to right) a Gaussian random signal (a,d,g), the data including noise (b,e,h), and the reconstructed map (c,f,i). The additive Gaussian white noise has a variance $\sigma_n^2$ that sets a signal-to-noise ratio $\left< \sigma_s\right>_\Omega/\sigma_n$ of roughly $2$. The same code has been applied to three different spaces (top to bottom), namely a 1D regular grid with $512$ pixels (a,b,c), a 2D regular grid with $256 \times 256$ pixels (d,e,f), and a \textsc{HEALPix} grid with $n_\mathrm{side} = 128$ corresponding to $196,608$ pixels on the $\mathcal{S}^2$ sphere (g,h,i). (All figures have been created by \textsc{NIFTy} using the \texttt{field.plot} method.)}
        \label{fig:wf}
    \end{figure*}

    An established and widely used inference algorithm is the Wiener filter~\citep{W49} whose implementation in \textsc{NIFTy} shall serve as a demonstration example.

    The underlying inference problem is the reconstruction of a signal, $\bb{s}$, from a data set, $\bb{d}$, that is the outcome of a measurement process \eqref{eq:linearmodel}, where the signal response, $\bb{R} \: \bb{s}$, is linear in the signal and the noise, $\bb{n}$, is additive. The statistical properties of signal and noise are both assumed to be Gaussian,
    \begin{align}
        \bb{s} &\curvearrowleft {\cal G}(\bb{s},\bb{S}) \propto \exp\left(-\tfrac{1}{2} \bb{s}^\T \bb{S}^{-1} \bb{s} \right)
        , \\
        \bb{n} &\curvearrowleft {\cal G}(\bb{n},\bb{N})
        .
    \end{align}
    Here, the signal and noise covariances, $\bb{S}$ and $\bb{N}$, are known \emph{a~priori}. The \emph{a~posteriori} solution for this inference problem can be found in the expectation value for the signal $\bb{m} = \left< \bb{s} \right>_{(\bb{s}|\bb{d})}$ weighted by the posterior $P(\bb{s}|\bb{d})$ . This map can be calculated with the Wiener filter equation,
    \begin{align}
        \label{eq:wf}
        \bb{m} = \underbrace{\left( \bb{S}^{-1} + \bb{R}^\T \bb{N}^{-1} \bb{R} \right)^{-1}}_{\bb{D}} \underbrace{\left( \bb{R}^\T \bb{N}^{-1} \bb{d} \right)}_{\bb{j}}
        ,
    \end{align}
    which is linear in the data. In the IFT framework, this scenario corresponds to a free theory as discussed in the work by \citet{EFK09}, where a derivation of Eq.~\eqref{eq:wf} can be found. In analogy to quantum field theory, the posterior covariance, $\bb{D}$, is referred to as the information propagator and the data dependent term, $\bb{j}$, as the information source.

    \begin{figure}[!t]
        \centering
        \includegraphics[width=0.5\textwidth]{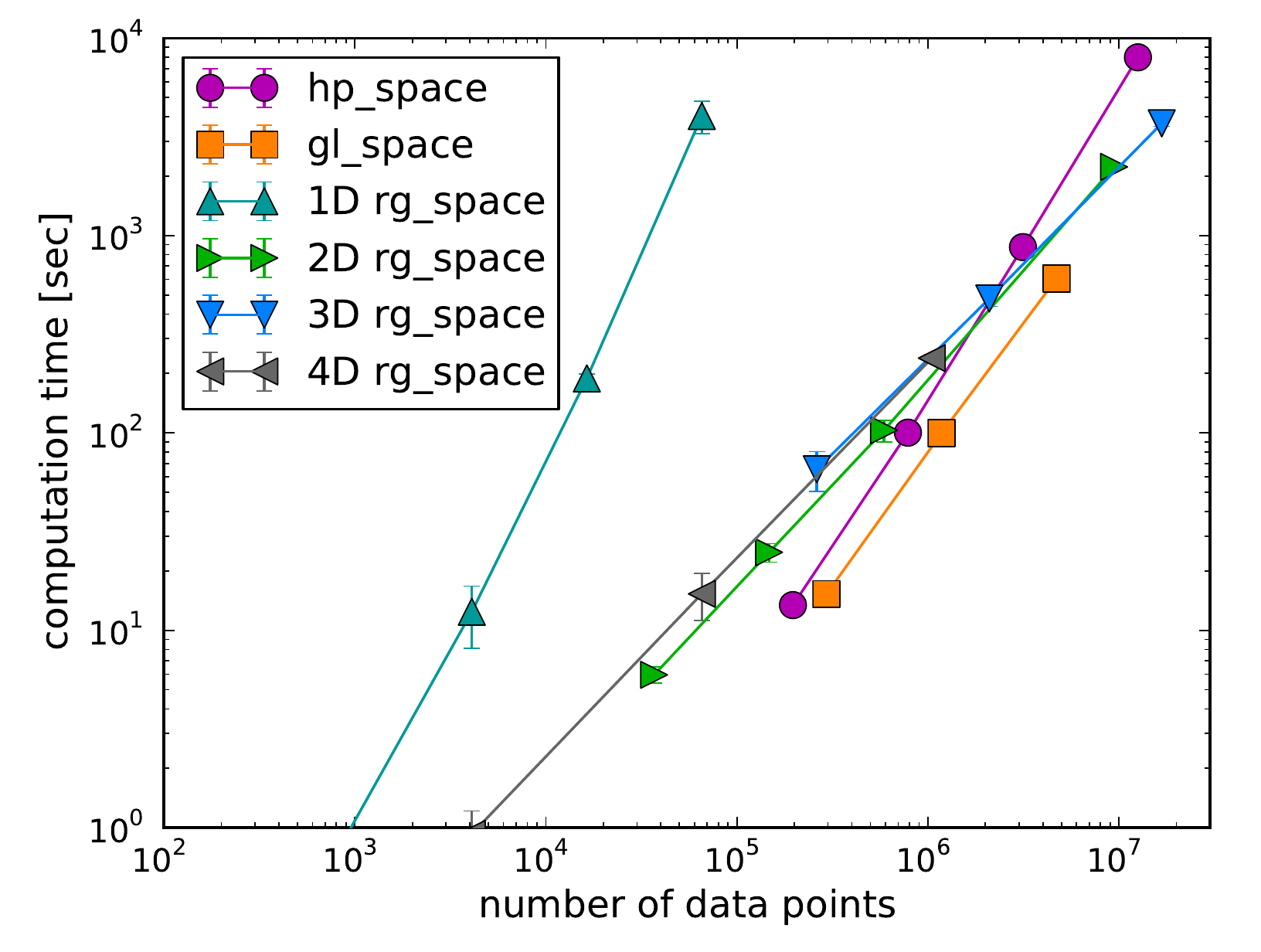}
        \flushleft
        \caption{Illustration of the performance of the Wiener filter code given in App.~\ref{sec:code} showing computation time against the size of the data set (ranging from $512$ to $256 \times 256 \times 256$ data points) for different signal spaces (see legend). The markers show the average runtime of multiple runs, and the error bars indicate their variation. (Related markers are solely connected to guide the eye.)}
        \label{fig:performance}
    \end{figure}

    The \textsc{NIFTy} based implementation is given in App.~\ref{sec:code}, where a unit response and noise covariance are used.\footnote{The Wiener filter demonstration is also part of the \textsc{NIFTy} package; see \texttt{nifty/demos/demo\_excaliwir.py} for an extended version.} This implementation is not only easily readable, but it also solves for $\bb{m}$ regardless of the chosen signal space; i.e., regardless of the underlying grid and its resolution. The functionality of the code for different signal spaces is illustrated in Fig.~\ref{fig:wf}.
    The performance of this implementation is exemplified in Fig.~\ref{fig:performance} for different signal spaces and sizes of data sets. A qualitative power law behavior is apparent, but the quantitative performance depends strongly on the used machine.

    \begin{figure*}[!t]
        \centering
        \begin{tabular}{ccc}
            \begin{overpic} [scale=0.3]{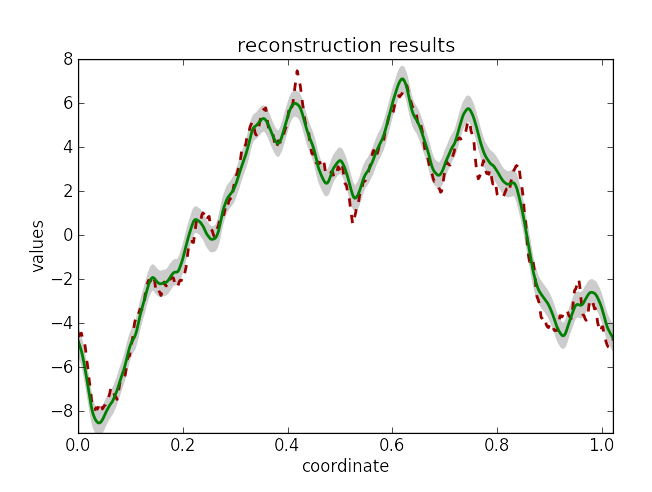} \put(-5,64){(a)} \end{overpic} &
            \begin{overpic} [scale=0.3]{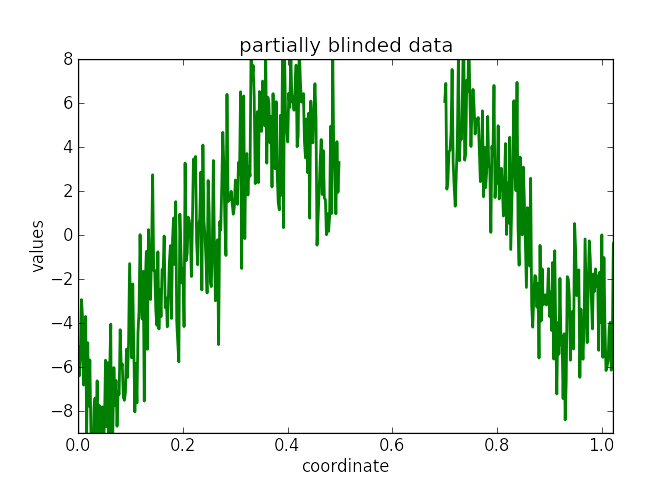} \put(-5,64){(b)} \end{overpic} &
            \begin{overpic} [scale=0.3]{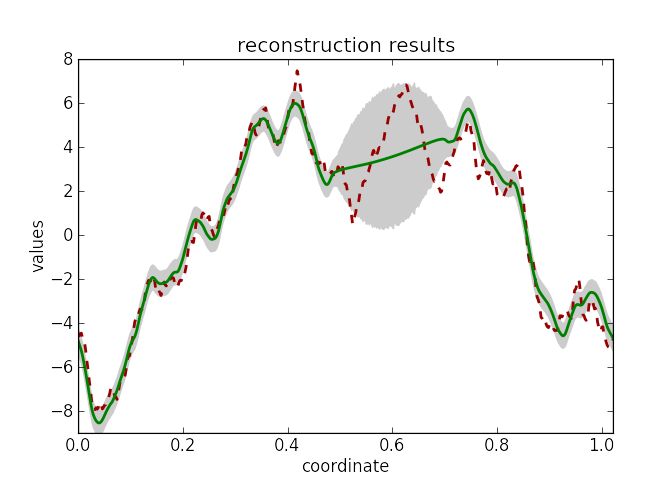} \put(-5,64){(c)} \end{overpic} \\
        \end{tabular}
        \flushleft
        \caption{Illustration of the 1D reconstruction results. Panel (a) summarizes the results from Fig.~\ref{fig:wf} by showing the original signal (red dashed line), the reconstructed map (green solid line), and the $1\sigma$-confidence interval (gray contour) obtained from the square root of the diagonal of the posterior covariance $\bb{D}$ that has been computed using probing; cf. Eq.~\eqref{eq:diag}. Panel (b) shows the 1D data set from Fig.~\ref{fig:wf} with a blinded region in the interval $[0.5,0.7]$. Panel (c) shows again the original signal (red, dashed line), the map reconstructed from the partially blinded data (green solid line), and the corresponding $1\sigma$-interval (gray contour) which is significantly enlarged in the blinded region indicating the uncertainty of the interpolation therein.}
        \label{fig:D}
    \end{figure*}

    The confidence in the quality of the reconstruction can be expressed in terms of a $1\sigma$-confidence interval that is related to the diagonal of $\bb{D}$ as follows,
    \begin{align}
        \sigma^{(\bb{m})} = \sqrt{\mathrm{diag}[\bb{D}]}
        .
    \end{align}
    The operator $\bb{D}$ defined in Eq.~\eqref{eq:wf} may involve inversions in different bases and thus is accessible explicitly only with major computational effort. However, its diagonal can be approximated efficiently by applying operator probing \eqref{eq:diag}. Fig.~\ref{fig:D} illustrates the 1D reconstruction results in order to visualize the estimates obtained with probing and to emphasize the importance of \emph{a~posteriori} uncertainties.

    \begin{figure*}[!t]
        \centering
        \begin{tabular}{ccc}
            \begin{overpic} [scale=0.3]{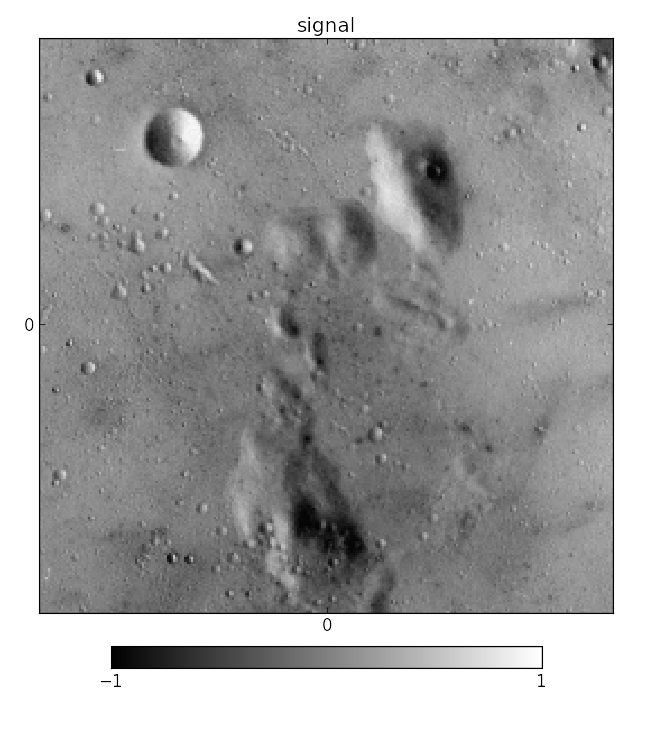} \put(-5,92){(a)} \end{overpic} &
            \begin{overpic} [scale=0.3]{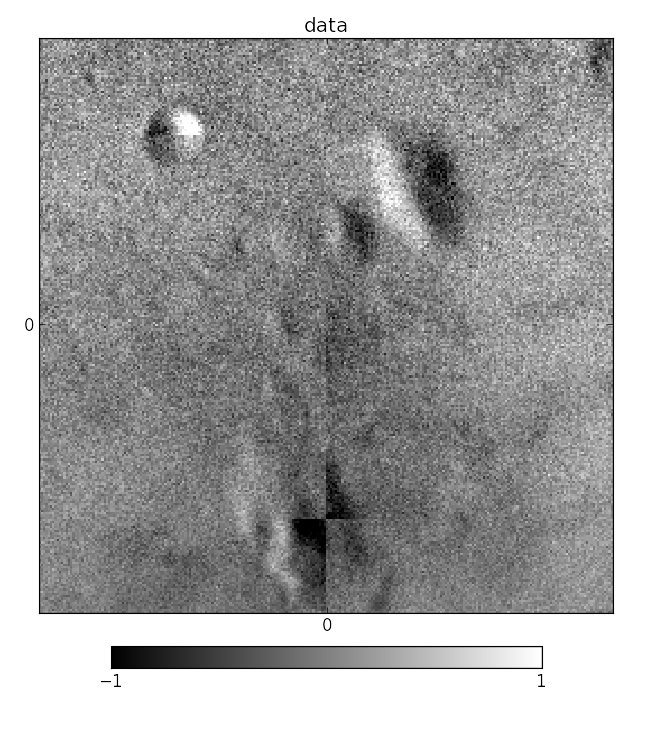} \put(-5,92){(b)} \end{overpic} &
            \begin{overpic} [scale=0.3]{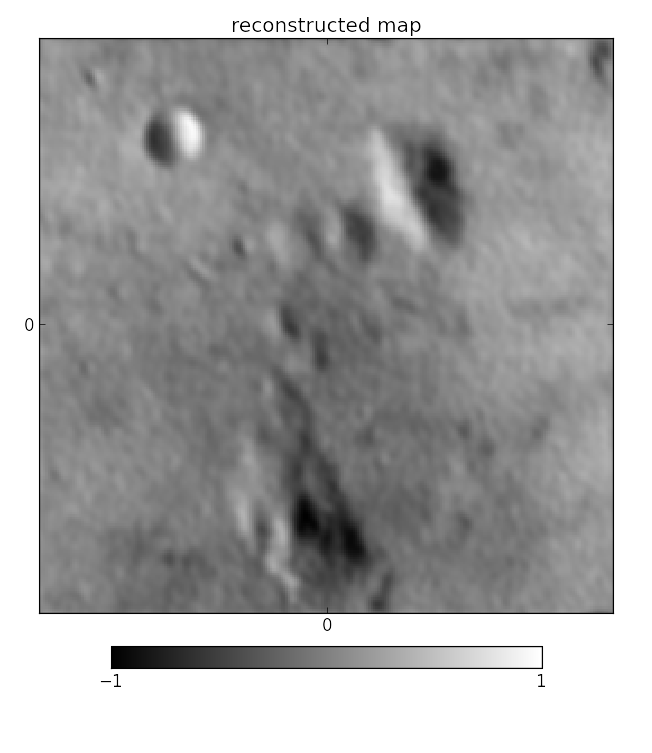} \put(-5,92){(c)} \end{overpic} \\
            \begin{overpic} [scale=0.3]{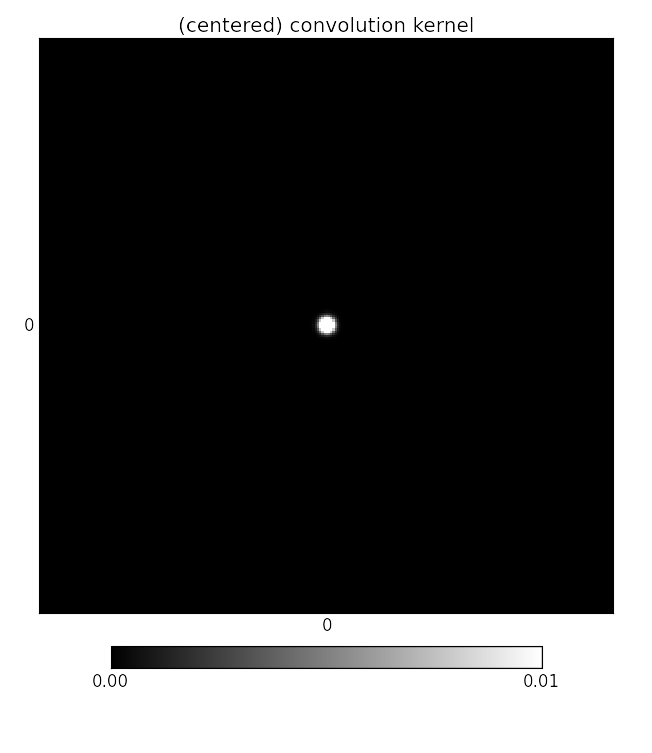} \put(-5,92){(d)} \end{overpic} &
            \begin{overpic} [scale=0.3]{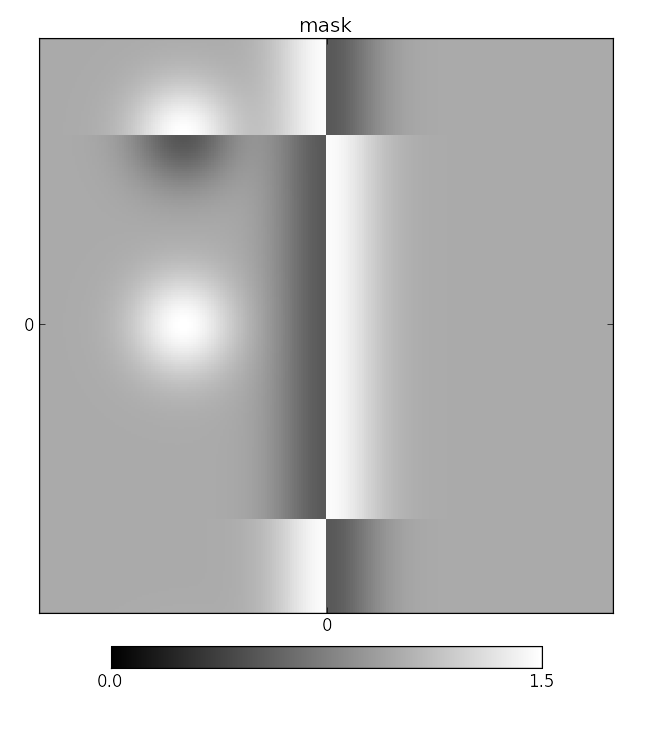} \put(-5,92){(e)} \end{overpic} &
            \begin{overpic} [scale=0.3]{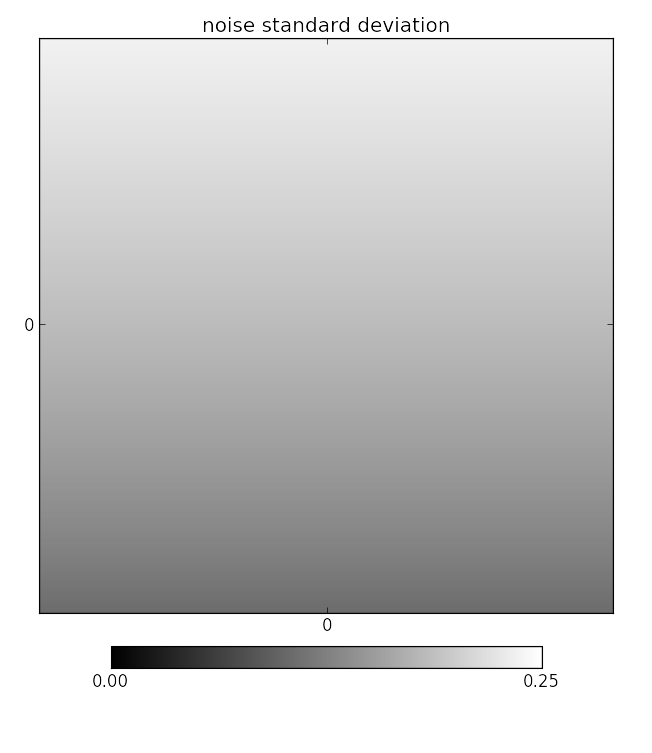} \put(-5,92){(f)} \end{overpic} \\
        \end{tabular}
        \flushleft
        \caption{Application of a Wiener filter to the classic ``Moon Surface'' image on a 2D regular grid with $256 \times 256$ pixels showing (top, left to right) the original ``Moon Surface'' signal (a), the data including noise (b), and the reconstructed map (c). The response operator involves a convolution with a Gaussian kernel (d) and a masking (e). The additive noise is Gaussian white noise with an inhomogeneous standard deviation (f) that approximates an overall signal-to-noise ratio $\left< \sigma_s\right>_\Omega/\left< \sigma_n\right>_\Omega$ of roughly $1$. (All figures have been created by \textsc{NIFTy} using the \texttt{field.plot} method.)}
        \label{fig:lena}
    \end{figure*}

    The Wiener filter code example given in App.~\ref{sec:code} can easily be modified to handle more complex inference problems. In Fig.~\ref{fig:lena}, this is demonstrated for the image reconstruction problem of the classic ``Moon Surface'' image\footnote{Source taken from the USC-SIPI image database at \url{http://sipi.usc.edu/database/}}. During the data generation~\eqref{eq:linearmodel}, the signal is convolved with a Gaussian kernel, multiplied with some structured mask, and finally, contaminated by inhomogeneous Gaussian noise. Despite these complications, the Wiener filter is able to recover most of the original signal field.

    \textsc{NIFTy} can also be applied to non-linear inference problems, as has been demonstrated in the reconstruction of log-normal fields with \emph{a~priori} unknown covariance and spectral smoothness \citep{OSBE12}. Further applications reconstructing three-dimensional maps from column densities \citep{prep1} and non-Gaussianity parameters from the cosmic microwave background \citep{prep2} are currently in preparation.

\section{Conclusions \& Summary}
\label{sec:conclusion}

    The \textsc{NIFTy} library enables the programming of grid and resolution independent algorithms. In particular for signal inference algorithms, where a continuous signal field is to be recovered, this freedom is desirable. This is achieved with an object-oriented infrastructure that comprises, among others, abstract classes for spaces, fields, and operators. \textsc{NIFTy} supports a consistent discretization scheme that preserves the continuum limit. Proper normalizations are applied automatically, which makes considerations by the user concerning this matter (almost) superfluous. \textsc{NIFTy} offers a straightforward transition from formulas to implemented algorithms thereby speeding up the development cycle. Inference algorithms that have been coded using \textsc{NIFTy} are reusable for similar inference problems even though the underlying geometrical space may differ.

    The application areas of \textsc{NIFTy} are widespread and include inference algorithms derived within both information field theory and other frameworks. The successful application of a Wiener filter to non-trivial inference problems illustrates the flexibility of \textsc{NIFTy}. The very same code runs successfully whether the signal domain is an $n$-dimensional regular or a spherical grid. Moreover, \textsc{NIFTy} has already been applied to the reconstruction of Gaussian and log-normal fields \citep{OSBE12}.


\section{Acknowledgments}

    We thank Philipp Wullstein, the \textsc{NIFTy} alpha tester Sebastian Dorn, and an anonymous referee for the insightful discussions and productive comments.

    Michael Bell is supported by the DFG Forschergruppe 1254 Magnetisation of Interstellar and Intergalactic Media: The Prospects of Low-Frequency Radio Observations.
    Martin Reinecke is supported by the German Aeronautics Center and Space Agency (DLR), under program 50-OP-0901, funded by the Federal Ministry of Economics and Technology.

    Some of the results in this paper have been derived using the HEALPix package \citep{G+05}. This research has made use of NASA's Astrophysics Data System.


\bibliography{NIFTY.bib}

\begin{appendix}
\section{Remark On Matrices}
\label{sec:matrices}

    The discretization of an operator that is defined on a continuum is a necessity for its computational implementation and is analogous to the discretization of fields; cf. Sec.~\ref{sec:discretization}. However, the involvement of volume weights can cause some confusion concerning the interpretation of the corresponding matrix elements. For example, the discretization of the continuous identity operator, which equals a $\delta$-distribution $\delta(x-y)$, yields a weighted Kronecker-Delta $\delta_{pq}$,
    \begin{align}
        \mathrm{id} \equiv \delta(x-y) \mapsto \big< \big< \delta(x-y) \big>_{\Omega_p} \big>_{\Omega_q} = \frac{\delta_{pq}}{V_q}
        ,
    \end{align}
    where $x \in \Omega_p$ and $y \in \Omega_q$. Say a field $\bb{\xi}$ is drawn from a zero-mean Gaussian with a covariance that equals the identity, $\mathcal{G}(\bb{\xi},\mathrm{id})$. The intuitive assumption that the field values of $\bb{\xi}$ have a variance of $1$ is not true. The variance is given by
    \begin{align}
        \left< \xi_p \xi_q \right>_{\{\bb{\xi}\}} = \frac{\delta_{pq}}{V_q}
        ,
    \end{align}
    and scales with the inverse of the volume $V_q$. Moreover, the identity operator is the result of the multiplication of any operator with its inverse, $\mathrm{id} = \bb{A}^{-1}\bb{A}$. It is trivial to show that, if $A(x,y) \mapsto A_{pq}$ and $\sum_q A_{pq}^{-1} A_{qr} = \delta_{pr}$, the inverse of $\bb{A}$ maps as follows,
    \begin{align}
        \bb{A}^{-1} \mapsto \big< \big< A^{-1}(x-y) \big>_{\Omega_p} \big>_{\Omega_q} = (A^{-1})_{pq} = \frac{A_{pq}^{-1}}{V_p V_q}
        ,
    \end{align}
    where $A_{pq}^{-1}$ in comparison to $(A^{-1})_{pq}$ is inversely weighted with the volumes $V_p$ and $V_q$.

    Since all those weightings are implemented in \textsc{NIFTy}, users need to concern themself with these subtleties only if they intend to extend the functionality of \textsc{NIFTy}.

\section{Libraries}

    \textsc{NIFTy} depends on a number of other libraries which are listed here for completeness and in order to give credit to the authors.
    \begin{itemize}
        \item[$\bullet$]
            \textsc{NumPy}, \textsc{SciPy}\footnote{\textsc{NumPy} and \textsc{SciPy} homepage \url{http://numpy.scipy.org/}} \citep{O06}, and several other \textsc{Python} standard libraries
        \item[$\bullet$]
            \textsc{GFFT}\footnote{\textsc{GFFT} homepage \url{https://github.com/mrbell/gfft}} for generalized fast Fourier transformations on regular and irregular grids; of which the latter are currently considered for implementation in a future version of \textsc{NIFTy}
        \item[$\bullet$]
            \textsc{HEALPy}\footnote{\textsc{HEALPy} homepage \url{https://github.com/healpy/healpy}} and \textsc{HEALPix} \citep{G+05} for spherical harmonic transformations on the \textsc{HEALPix} grid which are based on the \textsc{LibPSHT} \citep{R11} library or its recent successor \textsc{LibSHARP}\footnote{\textsc{LibSHARP} homepage \url{http://sourceforge.net/projects/libsharp/}} \citep{RS13}, respectively
        \item[$\bullet$]
            Another \textsc{Python} wrapper\footnote{libsharp-wrapper homepage \url{https://github.com/mselig/libsharp-wrapper}} for the performant \textsc{LibSHARP} library supporting further spherical pixelizations and the corresponding transformations
    \end{itemize}
    These libraries have been selected because they have either been established as standards or they are performant and fairly general.

    The addition of alternative numerical libraries is most easily done by the indroduction of new derivatives of the \texttt{space} class. Replacements of libraries that are already used in \textsc{NIFTy} are possible, but require detailed code knowledge.

\onecolumn
\section{Wiener Filter Code Example}
\label{sec:code}

    \footnotesize
    \begin{verbatim}


from nifty import *                                                   # version 0.3.0
from scipy.sparse.linalg import LinearOperator as lo
from scipy.sparse.linalg import cg


class propagator(operator):                                           # define propagator class

    _matvec = (lambda self, x: self.inverse_times(x).val.flatten())

    def _multiply(self, x):
        # some numerical invertion technique; here, conjugate gradient
        A = lo(shape=tuple(self.dim()), matvec=self._matvec, dtype=self.domain.datatype)
        b = x.val.flatten()
        x_, info = cg(A, b, M=None)
        return x_

    def _inverse_multiply(self, x):
        S, N, R = self.para
        return S.inverse_times(x) + R.adjoint_times(N.inverse_times(R.times(x)))

# some signal space; e.g., a one-dimensional regular grid
s_space = rg_space(512, zerocenter=False, dist=0.002)                 # define signal space
# or      rg_space([256, 256])
# or      hp_space(128)

k_space = s_space.get_codomain()                                      # get conjugate space
kindex, rho = k_space.get_power_index(irreducible=True)

# some power spectrum
power = [42 / (kk + 1) ** 3 for kk in kindex]

S = power_operator(k_space, spec=power)                               # define signal covariance
s = S.get_random_field(domain=s_space)                                # generate signal

R = response_operator(s_space, sigma=0.0, mask=1.0, assign=None)      # define response
d_space = R.target                                                    # get data space

# some noise variance; e.g., 1
N = diagonal_operator(d_space, diag=1, bare=True)                     # define noise covariance
n = N.get_random_field(domain=d_space)                                # generate noise

d = R(s) + n                                                          # compute data

j = R.adjoint_times(N.inverse_times(d))                               # define source
D = propagator(s_space, sym=True, imp=True, para=[S,N,R])             # define propagator

m = D(j)                                                              # reconstruct map

s.plot(title="signal")                                                # plot signal
d.cast_domain(s_space)
d.plot(title="data", vmin=s.val.min(), vmax=s.val.max())              # plot data
m.plot(title="reconstructed map", vmin=s.val.min(), vmax=s.val.max()) # plot map


    \end{verbatim}

\twocolumn

\end{appendix}
\end{document}